\documentclass[prb,showpaces,preprintnumbers,amsmath,amssymb,twocolumn,superscriptaddress,longbibliography]{revtex4-1}
\usepackage{graphicx}
\usepackage{dcolumn}
\usepackage{bm}
\usepackage{amssymb}
\usepackage{amsmath}
\usepackage{epstopdf}
\usepackage{booktabs}
\usepackage{threeparttable}
\usepackage[pdfstartview=FitH, CJKbookmarks=true, bookmarksnumbered=true, bookmarksopen=true, colorlinks, pdfborder=001, linkcolor=blue, anchorcolor=blue, citecolor=blue]{hyperref}
\begin{document}
\title{Magnetic field induced antiferromagnetic tricritical points in Ce$_2 $Sb and Ce$_2 $Bi}
\author{F. Wu}
\affiliation{Center for Correlated Matter and Department of Physics, Zhejiang University, Hangzhou 310058, China}
\author{C. Y. Guo}
\affiliation{Center for Correlated Matter and Department of Physics, Zhejiang University, Hangzhou 310058, China}
\author{Y. Chen}
\affiliation{Center for Correlated Matter and Department of Physics, Zhejiang University, Hangzhou 310058, China}
\author{H. Su}
\affiliation{Center for Correlated Matter and Department of Physics, Zhejiang University, Hangzhou 310058, China}
\author{A. Wang}
\affiliation{Center for Correlated Matter and Department of Physics, Zhejiang University, Hangzhou 310058, China}
\author{M. Smidman}
\email{msmidman@zju.edu.cn}
\affiliation{Center for Correlated Matter and Department of Physics, Zhejiang University, Hangzhou 310058, China}
\author{H. Q. Yuan}
\email{hqyuan@zju.edu.cn}
\affiliation{Center for Correlated Matter and Department of Physics, Zhejiang University, Hangzhou 310058, China}
\affiliation{Collaborative Innovation Center of Advanced Microstructures, Nanjing University, Nanjing 210093, China}

	\addcontentsline{toc}{chapter}{Abstract}

  \begin{abstract}
	 We present a detailed investigation of the physical properties of Ce$_2 $Sb and Ce$_2 $Bi single crystals, which undergo antiferromagnetic transitions at around 8.2 and 10~K respectively. When magnetic fields are applied parallel to the $c$~axis, metamagnetic transitions are observed at low temperatures, corresponding to a magnetic field-induced phase transition. It is found that the field-induced transition changes from second-order at higher temperatures, to first-order at low temperatures, suggesting the existence of tricritical points (TCPs) in both compounds. Since replacing Bi with Sb suppresses the TCP to lower temperatures and corresponds to a positive chemical pressure, these results suggest that applying pressure to Ce$_2$Sb may suppress the TCP to lower temperatures, potentially to zero temperature at a quantum tricritical point.
		
	\end{abstract}
	
	\maketitle
	
	\section{Introduction}		
Heavy fermion systems provide valuable opportunities for studying quantum phase transitions and quantum criticality. In many cases, long-range antiferromagnetic (AFM) magnetic order can be continuously suppressed to zero temperature at a quantum critical point (QCP) by tuning non-thermal parameters such as pressure, magnetic fields or doping \cite{GStewartRMP,gegenwart2008quantum,Weng2016,coleman2005quantum,ChenCPB}. In the vicinity of such a QCP, quantum critical phenomena and non-Fermi-liquid (NFL) behavior are commonly observed, which have often been explained on the basis of spin fluctuation theory \cite{coleman2005quantum,NFL,lohneysen2007}. However, the situation is different for ferromagnetic (FM) heavy fermions, where in itinerant FM systems a QCP is generally avoided \cite{Brando2016,Belitz1999}. One way for this to occur is instead of the ordering temperature being tuned continuously to zero with increasing pressure, there is a change in the ferromagnetic transition from second-order to first-order at a tricritical point (TCP) \cite{Brando2016,NFL}. When a field is applied, this leads to the characteristic tricritical wing structure in the temperature-pressure-field phase diagram, as found in UGe$_2$\cite{UGe2PRL} and ZrZn$_2$\cite{ZrZn2one,ZrZn2two}. Another means of avoiding a FM QCP is via a change of the magnetic ground state, as seen for instance in CeTiGe$_3$, where two new zero-field magnetic phases emerge at high pressures \cite{CeTiGe3}. This ultimately gives rise to a TCP at the termination of the first-order line separating the new magnetic ground state and the field-induced FM phase. On the other hand, the occurrence of TCPs in AFM systems is somewhat less common, although a high field study of USb$_2$ revealed a TCP between the AFM ground state and a field-induced ferrimagnetic phase \cite{USb2}. In some systems, the position of the TCP can be tuned by pressure or doping, which can allow for the continuous suppression of the TCP to zero temperature at a quantum tricritical point (QTCP). Such a scenario has been explored theoretically \cite{Takahiro2009,TCP2017PRL}, and recently the presence of a QTCP was experimentally identified in NbFe$ _2$ arising from the interplay of ferromagnetism and an intervening spin density wave phase \cite{friedemann2017quantum}.

The tetragonal compounds Ce$ _2 $Sb and Ce$ _2 $Bi are heavy fermion antiferromagnets, undergoing AFM transitions at around 8 and 10~K respectively \cite{isobe1987,canfield1991novel,oyamada1993heavy}. Both materials share the same layered structure, where there are two inequivalent Ce sites. Here there are significant differences in the nearest neighbor Ce-Ce distances between the two sites, where the separation between Ce(1) atoms is even shorter than in $\alpha$-Ce \cite{isobe1987}. Meanwhile, the crystalline-electric field (CEF) effect results in a strongly anisotropic magnetic susceptibility in Ce$_2$Bi \cite{canfield1991novel}, which is weaker in the case of Ce$_2$Sb \cite{oyamada1993heavy}. This may be due to differences in the wave function of the ground state doublet, which is consistent with the different splittings of the CEF levels found from neutron scattering \cite{Ohoyama19991189}.

Here we report detailed measurements of the physical properties of single crystals of Ce$ _2 $Sb and Ce$ _2 $Bi. Both materials exhibit AFM ground states, where in-field measurements reveal metamagnetic transitions which show a crossover from a second-order transition at elevated temperatures, to first-order behavior with clear hysteresis at low temperatures. These results indicate the presence of TCPs in the phase diagrams of these materials. Since Ce$_2$Sb corresponds to a positive chemical pressure relative to Ce$_2$Bi, and the TCP is at a lower temperature, this suggests that pressure may tune the TCP of these systems towards a zero temperature QTCP.

\section{Experimental Details}
Single crystals of Ce$_2 $Sb and Ce$_2 $Bi were grown using a Ce-self flux method with a molar ratio of Ce:Sb/Bi of 10:1, as described previously \cite{canfield1991novel}. The starting materials were placed in a Ta crucible which was sealed in an evacuated quartz tube, heated to 1175$^\circ$C before being cooled slowly to 900$^\circ$C, and centrifuged to remove excess Ce. The obtained crystals were plate-like, with typical dimensions 3mm$\times$3mm$\times$1mm. The composition of the crystals was checked using energy-dispersive x-ray spectroscopy. Resistivity and specific heat measurements were performed using a Quantum Design Physical Property Measurement System (PPMS) from 300~K to 1.8~K, with a maximum applied magnetic field of 9~T. The low temperature magnetoresistance was measured using a $^3$He system with a base temperature of 0.3~K. Resistivity measurements were performed after spot welding four Pt wires to the surface in the four probe geometry with an excitation current of 3~mA. For measurements of the dc susceptibility in applied fields up to 5~T and the ac susceptibility, a Quantum Design MPMS superconducting quantum interference device magnetometer was used. The coefficient of linear thermal expansion, $ \alpha = ({\rm d}l/{\rm d}T)/l$, where $ l $ is the sample length, was measured using a miniaturized high-resolution capacitance dilatometer \cite{Kuchler}. The sample was mounted inside the cell so that the length change of the sample along the $c$~axis was measured to a resolution higher than 0.1\AA, down to 0.3~K in an applied field.

\section{results}
\subsection{Antiferromagnetic transitions in Ce$_2 $Sb and Ce$_2 $Bi}

\begin{figure}[t]
	\begin{center}
		\includegraphics[width=\columnwidth]{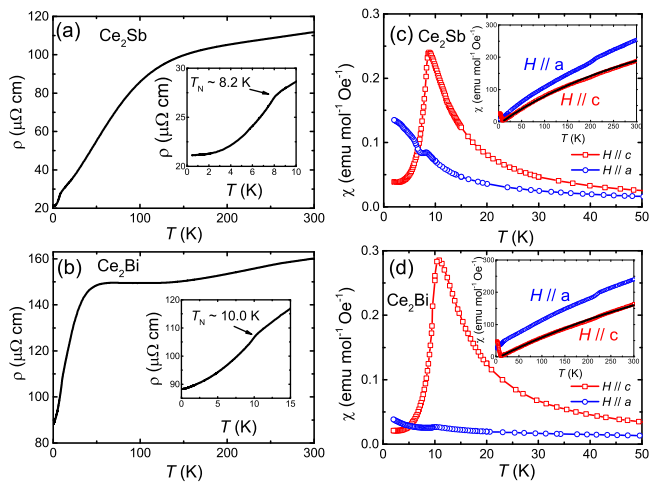}
	\end{center}
	\caption{Temperature dependence of the resistivity of (a) Ce$_2 $Sb, and (b) Ce$_2 $Bi between 0.3 and 300~K. The insets display the low temperature resistivity, which display anomalies at the antiferromagnetic transitions. Temperature dependence of the magnetic susceptibility of (c) Ce$_2 $Sb, and (d) Ce$_2 $Bi for applied fields of 0.01~T along the $a$ and $c$~axes. The insets display the inverse magnetic susceptibility up to 300~K, where the solid lines display the results from fitting with Curie-Weiss behavior. }
	\label{FIG1}
\end{figure}

\begin{figure}[t]
	\begin{center}
		\includegraphics[width=\columnwidth]{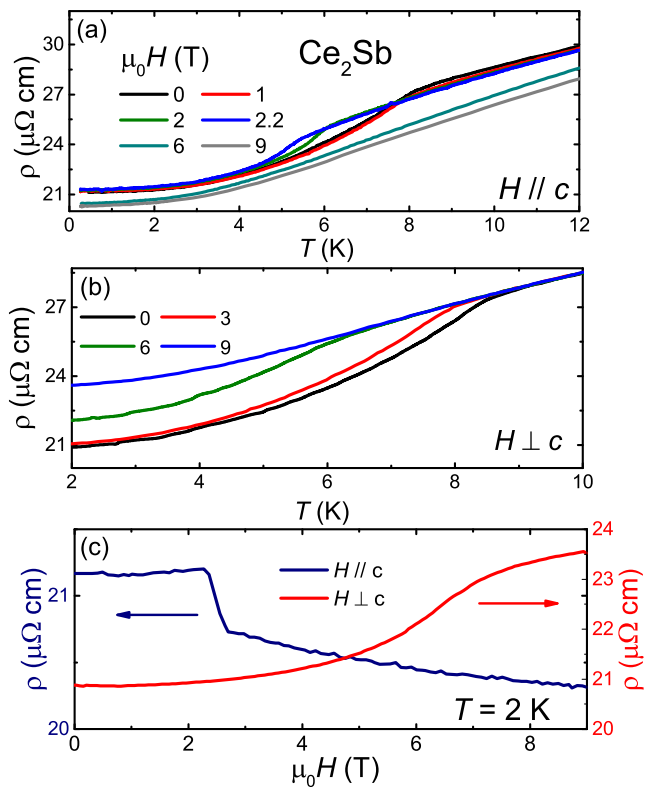}
	\end{center}
	\caption{Temperature dependence of the resistivity of Ce$_2 $Sb with different magnetic fields applied (a) parallel and (b) perpendicular to the easy $c$~axis, with the current along the $a$~axis. (c) Field dependence of the resistivity of Ce$_2 $Sb at 2~K, with applied fields both parallel and perpendicular to the $c$~axis.}
	\label{FIG2}
\end{figure}		
Figures~\ref{FIG1}(a) and (b) shows the temperature dependence of the resistivity ($\rho(T)$) of Ce$ _2 $Sb and Ce$ _2 $Bi between 0.3 and 300~K. Broad humps are observed in $\rho(T)$ of both compounds, which indicate the influence of the Kondo and CEF effects. The residual resistivity ratio (RRR=$\rho(300{\rm K})$/$\rho(2{\rm K})$) is about 5.3 for Ce$_2$Sb and 1.8 for Ce$_2$Bi. As shown in the insets, anomalies are observed at around 8.2~K and 10.0~K respectively, which correspond to the antiferromagnetic transitions reported previously \cite{isobe1987}. The temperature dependence of the magnetic susceptibility ($\chi(T)$) of the two materials is displayed in Figs.~\ref{FIG1}(c) and (d), for fields applied both along the $c$~axis and $a$~axis. The estimated demagnetization factors for both samples was around 0.59 for a field applied along the  $c$~axis, and upon correcting for this effect there was little change to $\chi(T)$. For both compounds, pronounced peaks are observed at the transition when the applied magnetic field is along the $c$~axis, below which $ \chi(T) $ decreases with decreasing temperature. This behavior is characteristic of a second-order antiferromagnetic transition, where the $c$~axis corresponds to the easy direction. Furthermore, $\chi(T)$ flattens at low temperatures, and the absence of a low temperature upturn suggests a lack of magnetic impurity phases in the crystalline samples. For fields applied along the hard $a$~axis, only small anomalies are observed at the transition, and $\chi(T)$ continues to increase below T$_N$, evidencing the significant anisotropy in these materials, which is greater in Ce$ _2 $Bi than Ce$ _2 $Sb. From analyzing $\chi(T)$ for $H\parallel c$ with the Curie-Weiss law at elevated temperatures, the estimated effective moments are about $ \mu_{\rm eff} $= 1.99~$ \mu_{\rm B} $/Ce and 2.35~$ \mu_{\rm B} $/Ce for Ce$ _2 $Sb and Ce$ _2 $Bi, while the respective Curie-Weiss temperatures are $-7$~K and $-8$~K. On the other hand, for fields applied along the $a$~axis, a small kink is observed at high temperatures for both compounds, which was previously reported for Ce$ _2 $Sb \cite{isobe1987}.
\begin{figure}[t]
	\begin{center}
		\includegraphics[width=\columnwidth]{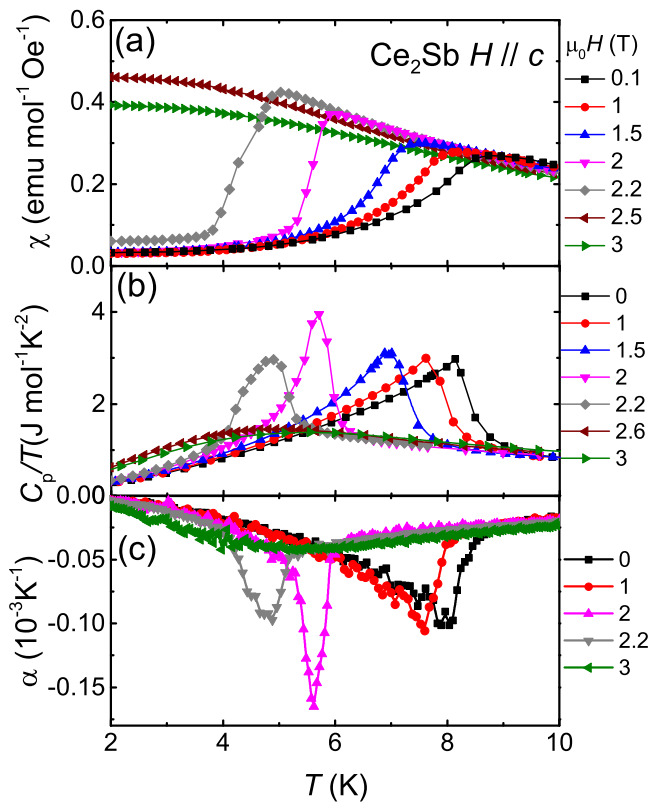}
	\end{center}
	\caption{(a) Temperature dependence of the (a) magnetic susceptibility, (b) specific heat, and (c) thermal expansion coefficient of Ce$_2 $Sb in various magnetic fields applied parallel to the $c$~axis.}
	\label{FIG3}
\end{figure}

Figure~\ref{FIG2}(a) displays the low temperature resistivity of Ce$_2 $Sb with different magnetic fields applied along the $c$~axis. With increasing magnetic field, the AFM transition is continuously suppressed from 8.2~K to lower temperatures, and for fields larger than 2.2~T along the $c$~axis, no magnetic transition can be observed. Meanwhile the resistivity when the field is applied perpendicular to the $c$~axis is displayed in Fig.~\ref{FIG2}(b), where the magnetic transition is suppressed more slowly with magnetic field. The field dependence of the resistivity of Ce$_2 $Sb at 2~K is shown in Fig.~\ref{FIG2}(c), where a metamagnetic transition is observed for fields applied along the $c$~axis onsetting at about 2.3~T, but no such transition is found up to 9~T for fields in the $ab$~plane. Such a metamagnetic transition where there is a sudden drop of the resistivity likely corresponds to a greater alignment of spins in-field along the easy axis direction. On the other hand, the gradual increase of $ \rho(H) $ for a field applied along the  $a$~axis  suggests that the spins do not significantly realign to be parallel to an in-plane field within this field range.

\begin{figure}[t]
	\begin{center}
		\includegraphics[width=0.9\columnwidth]{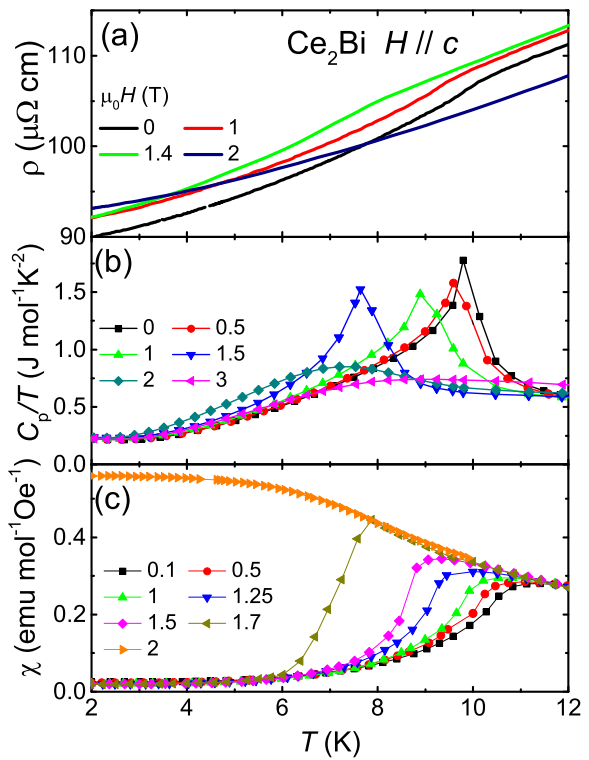}
	\end{center}
	\caption{Temperature dependence of the (a) resistivity, (b) specific heat, and (c) magnetic susceptibility of Ce$_2 $Bi in different magnetic fields applied parallel to the $c$~axis.}
	\label{FIG4}
\end{figure}

The temperature dependence of the magnetic susceptibility, heat capacity and thermal expansion coefficient $\alpha $ of Ce$_2 $Sb, are displayed in Fig.~\ref{FIG3}, for various magnetic fields applied parallel to the $c$~axis. The maxima in the magnetic susceptibility and specific heat correspond to the AFM transitions, which are shifted to lower temperature with increasing field, until at fields larger than 2.2~T no transitions can be observed down to 2~K. A broad Schottky-like hump in the specific heat is observed for fields above 2.2~T, which continuously moves to higher temperature with increasing field. Thermal expansion measurements reveal a negative $\alpha(T)$ at low temperatures, as is observed in several other heavy fermion systems \cite{URu2Si2}. It can be seen that upon decreasing the temperature, there is an abrupt decrease in $\alpha(T)$ at the antiferromagnetic transition, which is suppressed to lower temperature upon increasing the applied field. The rate of change of the $T_{\rm N}$ line in the $T-P$ phase diagram (${\rm d}T/{\rm d}P $) can be estimated using the Ehrenfest relation \cite{Ehrenfest,CaMnO3}, which is given by ${\rm d}T/{\rm d}P~=~TV\Delta\alpha/\Delta C_{\rm p} $. Here $\Delta\alpha$ and $ \Delta C_{\rm p} $ are the changes of $\alpha$ and $C_{\rm p}$ at $T_{\rm N}$, respectively, while $V$ is the molar volume. Using our zero-field data, $\Delta\alpha = -8\times $10$ ^{-5} $K$ ^{-1} $, and $ \Delta C_{\rm p} = 19$~J~mol$ ^{-1} $K$ ^{-1} $, which yields an estimated value of ${\rm d}T/{\rm d}P = -1.9$~K/GPa.

Figure~\ref{FIG4} shows the temperature dependence of the resistivity, specific heat and magnetic susceptibility of isostructural Ce$_2 $Bi. The AFM transition at $T_{\rm N}~=~10.0$~K for zero-field is suppressed to lower temperatures with increasing field along the $c$~axis, and for fields larger than 1.7~T as shown in Fig.~\ref{FIG4}(c), no magnetic transition is observed down to 2~K in all measurements. At higher fields, a broad hump is also observed in the specific heat, which moves to higher temperatures when the field is increased, similar to that observed in Ce$_2 $Sb.

\begin{figure}[t]
	\begin{center}
		\includegraphics[width=\columnwidth]{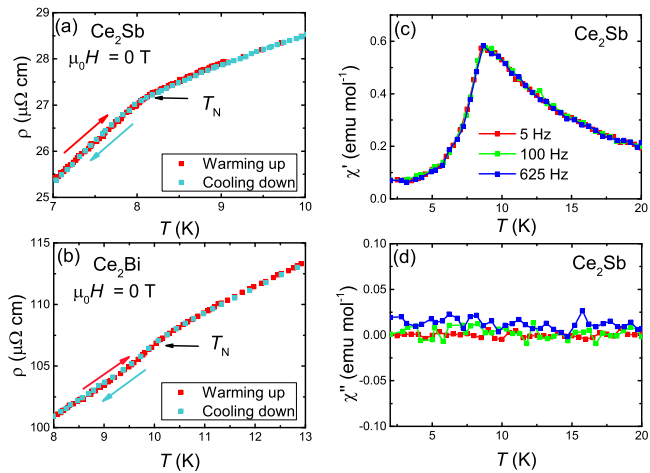}
	\end{center}
	\caption{Temperature dependence of the resistivity of (a) Ce$_2$Sb and (b) Ce$_2$Bi at temperatures close to $T_{\rm N}$, measured upon both warming up and cooling down. (c) Real part and (d) imaginary part of the ac susceptibility of Ce$_2$Sb with an excitation field of 0.25~mT along the $c$~axis. }
	\label{FIG5}
\end{figure}

Although the temperature dependence of $ \alpha $ for Ce$_2$Sb in zero-field displayed in Fig.~\ref{FIG3}(c) exhibits an anomaly at $ T_{\rm N} $, the continuous change through the transition suggests a second-order nature. This is further supported by the temperature dependence of the resistivity displayed in Figs.~\ref{FIG5} (a) and (b), as well as the ac susceptibility measurements in Figs.~\ref{FIG5} (c) and (d). Here there is no difference between the resistivity measurements performed upon warming and cooling through the transition, indicating that the AFM transitions for both compounds are second-order in zero-field. From the ac susceptibility measurements of Ce$_2$Sb, clear transitions are observed in the real part $\chi'$, where there is a sharp drop at the transition for an excitation field of 0.25~mT along the $c$~axis. Furthermore, the lack of frequency dependence of these results indicates the absence of spin-glass type behavior. Meanwhile the imaginary part of the susceptibility $\chi''$ is very small, showing temperature independent behavior with no anomaly at $T_{\rm N}$, providing further evidence for the second-order nature of the transition in zero-field. On the other hand, it can be seen in Fig.~\ref{FIG3}(b) that there is a distinct change in the temperature dependence of $C/T$ between 0~T and 2~T. Here the data in zero-field resembles a second-order transition, while in 2~T the anomaly is sharper and more $\delta$-like. This change suggests that the transition becomes first-order in higher fields, which is supported by the thermal expansion results in Fig.~\ref{FIG3}(c), where in a field of 2~T there is much more abrupt change of $\alpha$ at the magnetic transition, not observed in lower applied fields.

\subsection{Tricritical points in Ce$_2 $Sb and Ce$_2 $Bi under field}

\begin{figure}[t]
	\begin{center}
		\includegraphics[width=\columnwidth]{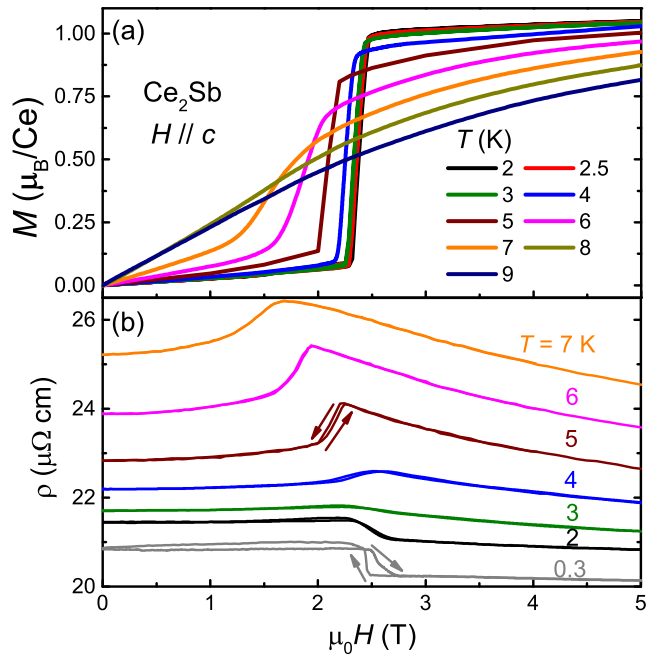}
	\end{center}
	\caption{(a) Magnetization of Ce$_2 $Sb as a function of applied magnetic field along the $c$~axis at several temperatures. (b) Magnetic field dependence of the resistivity of Ce$_2 $Sb at various temperatures for fields along the  $c$~axis .}
	\label{FIG6}
\end{figure}

Figure~\ref{FIG6} displays low temperature magnetization and magnetoresistance measurements of Ce$_2$Sb as a function of field along the $c$~axis at various temperatures. In both these measurements, abrupt metamagnetic transitions can be observed at low temperatures. The magnetization at 2~K exhibits a sharp jump at about 2.4~T with small hysteresis, above which there is little change of the magnetization upon further increasing the field up to 5~T, where it reaches a value of 1.05~$ \mu_{\rm B} $/Ce. Such a sharp change suggests that the field along the easy axis induces a sudden reorientation of the spins so that at low temperatures there is a first-order transition. Upon increasing the temperature, the transition moves to lower fields and becomes significantly broader, which is consistent with the field-induced suppression of the AFM phase field changing from first-order to second-order. The field dependence of the resistivity is displayed in Fig.~\ref{FIG6}(b). The metamagnetic transition is observed as an abrupt drop in the resistivity at low temperatures, where clear hysteresis can be observed between the up and down field sweeps. The size of the hysteresis loop decreases with increasing temperature, and is barely observable for $T >$ 5~K.

\begin{figure}[t]
	\begin{center}
		\includegraphics[width=\columnwidth]{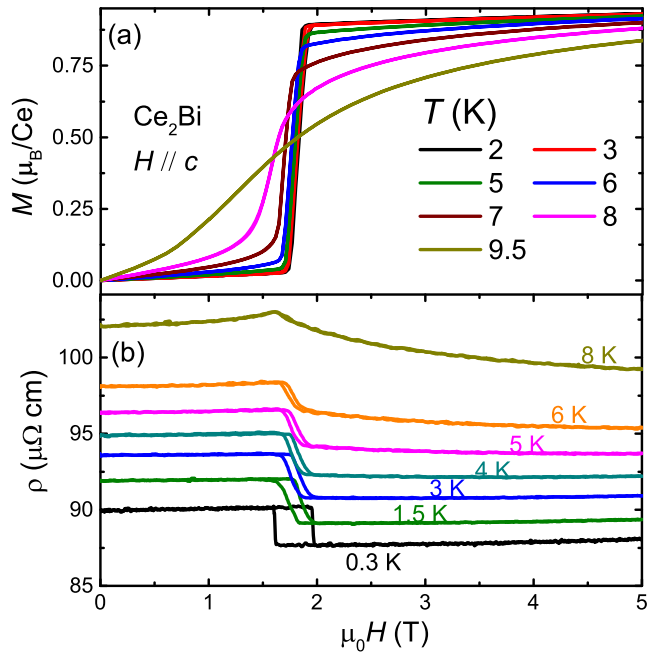}
	\end{center}
	\caption{Field dependence of the (a) magnetization, and (b) resistivity of Ce$_2 $Bi at various temperatures for magnetic fields applied along the $c$~axis.}
	\label{FIG7}
\end{figure}

\begin{figure}[t]
	\begin{center}
		\includegraphics[width=\columnwidth]{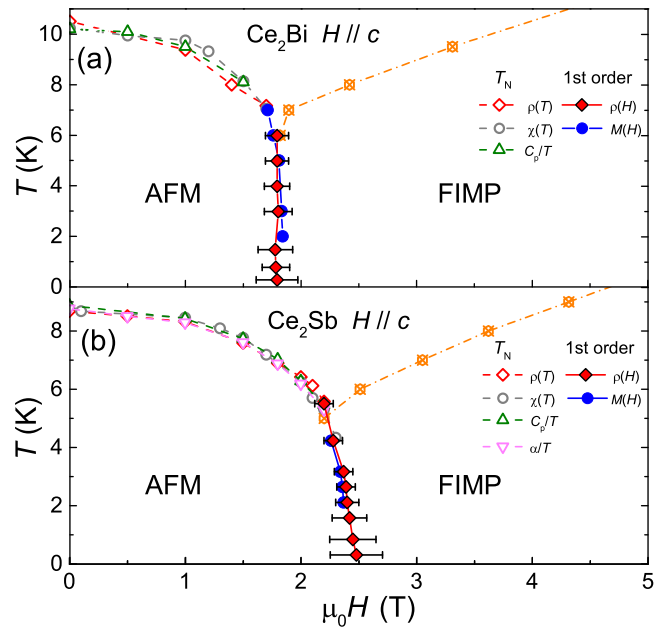}
	\end{center}
	\caption{Temperature-field phase diagram of (a) Ce$_2 $Bi, and (b) Ce$_2 $Sb, for fields along the $c$~axis. The hollow points show the positions of $T_{\rm N}$ determined from the temperature dependence of various quantities, while the solid symbols denote the first-order transition between the AFM phase and a field-induced magnetic phase (FIMP), where the error bars are obtained from the difference in the onset of the transition from measurements with up and down field sweeps. Upon lowering the temperature at higher fields, there is a crossover to the low temperature FIMP phase, where there is a larger magnetization. This crossover is marked by the dashed-dotted line, which corresponds to where the magnetization reaches 0.75$M_{\rm 2K,5T} $ (Table.~\ref{table}).}
	\label{FIG8}
\end{figure}

A similar field-induced transition can be observed in Ce$_2 $Bi, as shown in Fig.~\ref{FIG7}(a). At 2~K there is a very sharp metamagnetic transition at around 1.7~T, where there is a jump of the magnetization, reaching 0.935~$\mu_{\rm B}$/Ce at 5~T. The metamagnetic transition is also observed in the magnetoresistivity measurements displayed in Fig.~\ref{FIG7}(b). The field dependence of the resistivity of Ce$_2 $Bi at 0.3~K exhibits a significant drop at the metamagnetic transition, and there is also clear hysteresis between up and down field sweeps. With increasing temperature the transition broadens, and the hysteresis becomes smaller, and disappears above 6~K. This suggests that similar to Ce$_2 $Sb, the transition out of the AFM state is second-order at higher temperatures, but first-order at low temperatures. The nature of the field-induced state above the metamagnetic transition remains to be determined, in particular whether this corresponds to a field-induced spin-polarized state, or a different field-induced magnetic state with a net magnetization less than the saturated value. At low temperatures, the magnetization at 5~T corresponds to 1.05~$ \mu_{\rm B}$/Ce and 0.935~$ \mu_{\rm B}$/Ce for Ce$_2 $Sb and Ce$_2 $Bi respectively. In the case of USb$_2$, it could be determined that the jump in magnetization at the metamagnetic transition corresponded to just half of the saturated value from a comparison with neutron diffraction results \cite{USb2,USb2Neut}, which indicates that the spins are not fully aligned in the high field state. However, for Ce$_2 $Sb and Ce$_2 $Bi, we are not able to determine if the magnetization at 5~T corresponds to the full value of the Ce moments and therefore whether the system is in the spin polarized state . While neutron diffraction measurements have been performed on Ce$_2 $Sb, the value of the ordered moment is not reported \cite{isobe1987}. Furthermore, the CEF scheme is difficult to determine from analyzing $\chi(T)$, since there are two Ce sites, one with tetragonal and another with orthorhombic symmetry, and therefore such an analysis would have a large number of free parameters. On the other hand, the temperature dependent specific heat in high fields along the $c$~axis for both compounds exhibits a broad peak rather than a clear magnetic transition, suggesting that this corresponds to a crossover to the spin polarized state, but direct probes of the ordered moment such as neutron diffraction, or a determination of CEF level scheme, are desirable to confirm this.

Temperature-field phase diagrams for Ce$_2 $Bi and Ce$_2 $Sb are displayed in Fig.~\ref{FIG8}(a) and (b) respectively for fields along the $c$~axis, based on resistivity, magnetization, specific heat and thermal expansion measurements. The solid symbols show the first-order transition line determined from magnetoresistance and field-dependent magnetization measurements. The error bars are based on the differences in the fields where the metamagnetic transitions occur between up-sweep and down-sweep measurements. The  dashed-dotted line marks the crossover to the low temperature field-induced phase, where the magnetization is larger. Since the field-induced transition is second-order at higher temperatures, and first-order at low temperatures, this suggests the presence of a TCP at the point where the transition changes from second to first-order.

\begin{figure}[t]
	\begin{center}
		\includegraphics[width=\columnwidth]{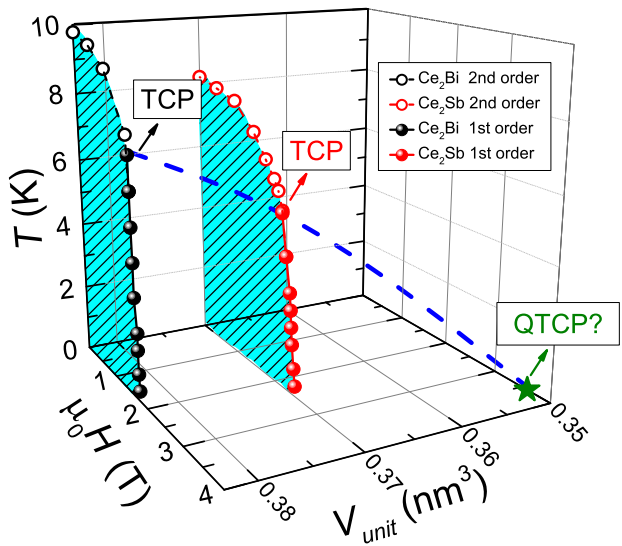}
	\end{center}
	\caption{Illustration of the effects of pressure on Ce$_2 $Bi and Ce$_2 $Sb. The black and red data points denote the antiferromagnetic transition temperatures of Ce$_2 $Bi and Ce$_2 $Sb respectively. As the unit cell volume is reduced, both $T_{\rm N}$ and the TCP shift to lower temperatures. This suggests that the further application of pressure may suppress the TCP to zero temperature at a quantum tricritical point.}
	\label{FIG9}
\end{figure}

A comparison between the properties of Ce$_2 $Sb and Ce$_2 $Bi is displayed in Table.~\ref{table}. The replacement of Bi with Sb corresponds to a positive chemical pressure effect, where the unit cell volume is reduced but the crystal structure remains unchanged. The effect of this positive pressure is to suppress the AFM transition temperature while increasing the critical field between the AFM and field-induced phases ($B _c$). It has been proposed that in Ce-based materials the magnetic phase will eventually be suppressed by pressure\cite{Hitoshi2016}, and the TCP moves to lower temperature with pressure, eventually reaching a QTCP \cite{Takahiro2009}. As shown in Fig.~\ref{FIG9}, Ce$_2 $Sb corresponds to a higher chemical pressure than Ce$_2 $Bi, and correspondingly the TCP is found at lower temperature. This suggests that a QTCP may be realized in this system upon applying a sufficiently large pressure. 

The estimated value of ${\rm d}T/{\rm d}P $ for $ T_{\rm N} $ using the Ehrenfest relation is about -1.9~GPa/K, which would suggest that the antiferromagnetic phase of Ce$_2 $Sb can be suppressed by applying a pressure of around 4~GPa. This value is comparable to other Ce-based heavy fermion compounds with similar magnetic transition temperatures, such as CePd$ _2 $Si$ _2 $, which has an antiferromagnetic transition at 10~K that is suppressed to zero temperature at around 2.5 GPa \cite{CePd2Si2}. On the other hand, if the TCP were suppressed at a similar rate with pressure as $T_{\rm N}$, then the QTCP would be reached at a lower pressure of 3~GPa. The realization of a QTCP in a Ce-based system has been reported in the temperature-pressure-magnetic field phase diagram of CeTiGe$ _3 $ \cite{CeTiGe3}. This compound orders ferromagnetically at $ T_{\rm C} $ = 14~K, but under pressure there is a change of magnetic ground state, where two likely AFM transitions are observed above 4.1~GPa. TCPs are found in the $ T$-$H $ phase diagrams at pressures above 4.1~GPa, at the point where the transition to the field-induced FM state changes from first to second-order. The TCP line is suppressed with pressure, which is extrapolated to a QTCP at 5.4~GPa and 2.8~T. On the other hand, CeAuSb$ _2 $ exhibits AFM order below $ T_{\rm N} $ = 6~K, where the transition from the AFM to paramagnetic state was found be first-order at low temperatures \cite{CeAuSb2}, and two critical end points were identified in the $ T$-$H $ phase diagram \cite{CeAuSb3}. However, while the AFM transition is initially suppressed by pressure, this phase is replaced by a different magnetic ground state above 2.1~GPa, and the fate of the in-field first-order transitions under pressure is not reported \cite{CeAuSb2pressure}. As such it remains an open question whether pressure can smoothly suppress the TCP of Ce$_2 $Sb to zero temperature at a QTCP as identified in CeTiGe$_3$, or if as in the case of CeAuSb$ _2 $, there is a change of magnetic ground state.

\begin{table}
	\centering
	\caption{Comparison of various parameters between Ce$_2 $Sb and Ce$_2 $Bi. The two compounds are isostructural, but with slightly different lattice parameters and unit cell volumes ($V_{\rm unit}$). The antiferromagnetic ordering temperature ($T_{\rm N}$), critical field of the AFM phase at 0.3~K ($B_{\rm c}$), and magnetization in the field-induced phase at 2~K ($M_{\rm 2K,5T} $) are all also displayed.}
	\begin{tabular}{p{2.5cm}|p{2.5cm}|p{2.5cm}}
		\hline
		\hline
		& Ce$_2$Bi & Ce$_2$Sb  \\
		\hline
		$ V_{\rm unit}$ (nm$ ^3 $) & 0.383  & 0.370  \\
		\hline
		Space group & $I4/mmm$ & $I4/mmm$  \\
		\hline
		$ T_{\rm N}  $(K)& 10.0 & 8.2  \\
		\hline
		$ B_{\rm c} $ (T)& 1.7 & 2.5  \\
		\hline
		$M_{\rm 2K,5T} $($ \mu_{\rm B}$/Ce )& 0.935 & 1.05  \\
		\hline
		\hline
	\end{tabular}
	
	\label{table}
\end{table}

\section{Conclusion}
We show evidence for the existence of tricritical points in the field-temperature phase diagrams of Ce$_2 $Sb and Ce$_2 $Bi from transport, magnetic and thermodynamic measurements of single crystals. For fields applied along the $c$~axis, a clear metamagnetic transition is observed below $T _N $. This suppression of the AFM state by magnetic field is second-order at higher temperatures but becomes first-order at low temperatures, suggesting the existence of TCPs in both compounds. Since the replacement of Bi with Sb corresponds to a positive chemical pressure, the effect of pressure is to shift the TCP to lower temperatures. Consequently, the application of hydrostatic pressure to Ce$_2$Sb may suppress the TCP to zero temperature, allowing for the realization of a QTCP.

\begin{acknowledgments}

We thank X.~Lu and F.~Steglich for interesting discussions and helpful suggestions. This work was supported by the National Key R\&D Program of China
(No.~2017YFA0303100 and No.~2016YFA0300202), the National Natural Science Foundation
of China (No. U1632275, No. 11604291), and the Science Challenge Project of China
(No.~TZ2016004).

\end{acknowledgments}

\end{document}